\documentclass[10pt,conference]{IEEEtran}\IEEEoverridecommandlockouts
\usepackage{epsfig,graphics,subfigure,psfrag,amsmath,cases}
\usepackage{latexsym,amssymb,amsmath,epsfig,subfigure,algorithm,amsthm}
\usepackage{algorithmic}
\usepackage{color}
\usepackage{url}
\usepackage{scrtime}
\usepackage{bm}
\usepackage{graphicx}
\usepackage{epstopdf}

\author{\vspace*{-0.5mm} \IEEEauthorblockN{Elena Boshkovska\IEEEauthorrefmark{1}, Rania Morsi\IEEEauthorrefmark{1}, Derrick Wing Kwan Ng\IEEEauthorrefmark{2}, and Robert Schober\IEEEauthorrefmark{1}
\thanks{Robert Schober is also with the University of British Columbia. This work was supported in part by the AvH Professorship Program of the Alexander von Humboldt Foundation.}
} \vspace*{0.5mm}
\IEEEauthorblockA{\IEEEauthorrefmark{1}Friedrich-Alexander-University Erlangen-N\"urnberg (FAU), Germany\\} \vspace*{0.5mm}
 \IEEEauthorblockA{\IEEEauthorrefmark{2}The University of New South Wales, Australia
    \\}
\vspace*{-11.5mm}

}

\title{Power Allocation and Scheduling for SWIPT Systems with Non-linear Energy Harvesting Model}

\date{\thistime,\,\today}

\newtheorem{Thm}{Theorem}

\newcommand{\norm}[1]{\lVert#1\rVert}

\DeclareMathOperator{\maxo}{\mathrm{maximize}}

\begin{document}

\maketitle

\begin{abstract}

In this paper, we design a resource allocation algorithm for multiuser simultaneous wireless information and power transfer systems for a realistic non-linear energy harvesting (EH) model.
In particular, the algorithm design is formulated as a non-convex optimization problem for the maximization of the long-term average total harvested power at EH receivers subject to quality of service requirements for information decoding receivers. To obtain a tractable solution, we transform the corresponding non-convex sum-of-ratios objective function into an equivalent objective function in parametric subtractive form. This leads to a computationally efficient iterative resource allocation algorithm. Numerical results reveal a significant performance gain that can be achieved if the resource allocation algorithm design is based on the non-linear EH model instead of the traditional linear model.
\end{abstract}
\vspace*{1mm}

\renewcommand{\baselinestretch}{0.98}
\large\normalsize

\section{Introduction}
\vspace*{-1mm}
\label{sect1}
Energy harvesting (EH) is an appealing solution for enabling self-sustainable wireless devices in communication networks. Thereby, the inconvenience of recharging and replacing batteries can be avoided by harvesting energy from different energy sources, such as solar and wind. Recently, wireless power transfer (WPT) via radio frequency (RF) signals has received considerable attention as it provides an ubiquitous, relatively stable, and controllable source of energy \cite{Krikidis2014,CN:Shannon_meets_tesla}. Moreover, additional benefits can be reaped by employing information-carrying signals for WPT, which enables simultaneous wireless information and power transfer (SWIPT) \cite{CN:WIPT_fundamental}.

SWIPT introduces a paradigm shift for system, receiver, and resource allocation algorithm design for communication systems due to the newly imposed challenges in delivering information and energy concurrently. In particular, there is a fundamental trade-off between EH and information decoding (ID), as was shown in \cite{CN:WIPT_fundamental}. Thereby, resource allocation plays a particularly important role for improving the system performance of SWIPT networks. In \cite{CN:Shannon_meets_tesla}, the authors proposed a power allocation algorithm for near-field communication systems. However, the authors of \cite{CN:Shannon_meets_tesla} and \cite{CN:WIPT_fundamental} assumed that the receivers are able to harvest energy from the received signal, while simultaneously decoding the embedded information, which is not feasible in practice, yet, due to practical limitations of EH circuits. Consequently, hybrid power splitting receivers and separate receivers were proposed for SWIPT in \cite{CN:WIP_receiver} and \cite{CN:MIMO_WIPT}, respectively. Additionally, a simple time-switching receiver was proposed for alternating between ID and EH across different time slots \cite{CN:WIP_receiver}.
For multiuser downlink SWIPT systems, suboptimal order-based scheduling schemes to balance the trade-off between the ergodic achievable rates and the average amounts of harvested energy of the users were proposed in \cite{multiuser_scheduling}. Furthermore, adopting the same system model as in \cite{multiuser_scheduling}, optimal multiuser scheduling schemes guaranteeing a long-term minimum harvested energy for SWIPT were reported in \cite{CN:Maryna_2015}.

The fundamental element of SWIPT systems that enables RF-EH is the EH circuit. The EH circuit includes a rectifier as the component that converts the power of the received RF signal to direct current (DC) power with a certain conversion efficiency \cite{valenta2014harvesting}. On the other hand, the design of resource allocation algorithms in SWIPT systems relies on an accurate mathematical model for the characteristics of the EH circuit implemented at the EH receiver. For instance, practically all existing works, e.g. \cite{Krikidis2014}, \cite{CN:WIP_receiver}--\nocite{CN:MIMO_WIPT,multiuser_scheduling}\cite{CN:Maryna_2015}, assume a specific linear EH model for the RF-to-DC power conversion.
However, practical EH circuits usually result in non-linear end-to-end WPT \cite{valenta2014harvesting}--\nocite{le2008efficient}\cite{guo2012improved}. Therefore, the traditional linear EH model adopted in the literature for resource allocation algorithm design may not be able to capture the non-linear characteristics of the RF-to-DC power conversion in practical RF-EH systems. Recently, a practical non-linear EH model was proposed in \cite{comm_letter:practical}, along with a beamforming algorithm for a downlink multi-antenna SWIPT system serving multiple information receivers (IRs) and multiple energy harvesting receivers (ERs). Specifically, the beamforming algorithm in \cite{comm_letter:practical} was designed for short-term maximization of the total harvested power at the ERs, while guaranteeing minimum required signal-to-interference-plus-noise ratios (SINRs) at multiple IRs. The results in \cite{comm_letter:practical} revealed that resource allocation algorithms designed for the simple linear EH model, which is widely used in the literature, may lead to resource allocation mismatches for practical non-linear EH circuits. However, the problem of joint user scheduling and long-term power allocation for SWIPT systems with practical non-linear EH circuits has not been considered in the literature, yet. Although scheduling schemes that exploit multiuser diversity for improving the performance of multiuser SWIPT systems were studied in \cite{multiuser_scheduling,CN:Maryna_2015}, the authors adopted the existing linear EH model for the end-to-end WPT, which may lead to suboptimal performance in practice.

In this paper, we adopt the practical non-linear EH model from \cite{comm_letter:practical} and study the optimal algorithm design for joint user scheduling and long-term power allocation in order to facilitate EH and to exploit multiuser diversity in multiuser downlink SWIPT systems. Thereby, the resource allocation algorithm design is formulated as a non-convex optimization problem with a sum-of-ratios objective function. Exploiting a recent result from the mathematical literature \cite{jonga2012efficient}, this difficult non-convex problem is solved optimally by a computationally efficient iterative algorithm after transforming the sum-of-ratios objective function into an equivalent objective function in subtractive form. Simulation results reveal significant improvements in performance, when the non-linear EH model is adopted for resource allocation algorithm design instead of the simple linear EH model.
\section{System Model}
\label{sect:network_model}
In this section, we present the adopted channel model and introduce the considered non-linear EH model.
\subsection{Channel Model}
\begin{figure}
\centering
\includegraphics[width= 2.99in, height=2in]{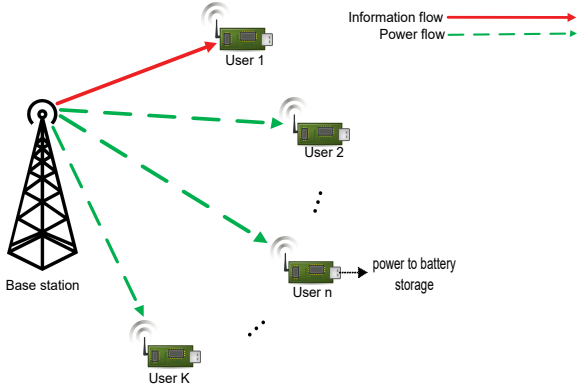}
\vspace*{-1mm}
\caption{A downlink multiuser SWIPT communication system with $K$ users.}
\vspace*{-2mm}
\label{fig:system_model}
\end{figure}

We consider a downlink multiuser system, where a single-antenna base station broadcasts the RF signal to $K$ single-antenna receivers capable of ID and EH, cf. Figure~\ref{fig:system_model}. We note that the receivers may also exploit other energy sources such that their power supply does not solely rely on the power harvested through RF-EH. Transmission in the system is divided into $T$ orthogonal time slots. In every time slot $n \in \{1,\ldots,T\}$, we perform joint user scheduling and power allocation to optimize the system performance. We assume a frequency flat slow fading channel. The downlink received symbol at user $k \in \{1,\ldots,K\}$ in time slot $n$ is given by
\begin{equation}
\label{eq_1}
y_k(n) = \sqrt{P_k(n) h_k(n)}x_k(n) + z_k(n),
\end{equation}
where $x_k(n)$ is the transmitted symbol, $P_k(n)$ is the transmit power, and $h_k(n)$ is the channel gain coefficient including the joint effects of multipath fading and path loss for user $k$ in time slot $n$. For the transmitted symbol, we assume a zero mean symbol with unit variance, i.e.,  $\mathbb{E}\{|x_k(n)|^2\} = 1, \forall n,k$, where $\mathbb{E}\{\cdot\}$ stands for statistical expectation. Furthermore, $z_k(n)$ represents the additive white Gaussian noise (AWGN) in time slot $n$ at user $k$ with zero mean and variance $\sigma^2$. Assuming perfect channel state information (CSI) at the user, the maximum achievable data rate (bit/s/Hz), i.e., the instantaneous capacity, for user $k$ in time slot $n$ is given by
\begin{equation}\label{eq_2}
C_k(n) = \log_2 \bigg( 1 + \frac{P_k(n)h_k(n)}{\sigma^2}\bigg).
\end{equation}
In each time slot, a single receiver is selected as the IR. Exploiting the broadcast nature of the wireless channel, the remaining $K-1$ receivers are scheduled as ERs to opportunistically harvest RF energy.

\subsection{Energy Harvesting Model}
\label{sect:receiver}
\begin{figure}
\centering
\includegraphics[width=2.95in, height=1.2in]{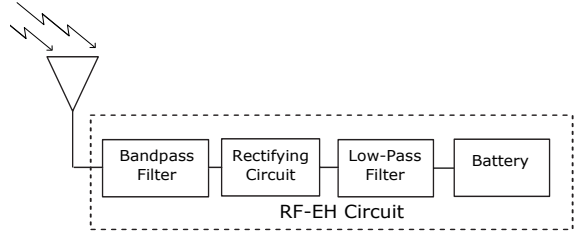}
\vspace*{-1mm}
\caption{Block diagram of an end-to-end RF-EH model.}
\vspace*{-2mm}
\label{fig:wpt_system}
\end{figure}

\begin{figure}
\centering
\includegraphics[width= 3.5in]{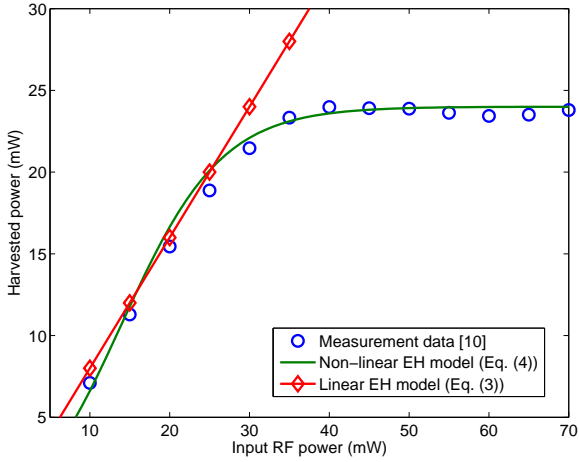}
\vspace*{-1mm}
\caption{Comparison between measurement data from \cite{guo2012improved}, the harvested power for the non-linear EH model in \eqref{eq_4}, and the linear EH model in \eqref{eq_3}.}
\vspace*{-2mm}
\label{fig:curve_fit}
\end{figure}

Figure~\ref{fig:wpt_system} depicts the EH receiver part of a general SWIPT system. In general, the RF-EH circuit consists of a bandpass filter, a rectifying circuit, and a low-pass filter followed by a battery \cite{valenta2014harvesting}. The bandpass and low-pass filters perform passive filtering of the signal in order to achieve impedance matching and removal of high frequency harmonic components, respectively. After bandpass filtering, the RF signal is rectified yielding DC power as output. The harvested energy at the ER is typically modelled based on a linear energy harvesting model \cite{Krikidis2014}, \cite{CN:WIP_receiver}--\nocite{CN:MIMO_WIPT,multiuser_scheduling}\cite{CN:Maryna_2015}:
\begin{equation}\label{eq_3}
P_{\text{ER-DC}}^{\text{linear}}=\eta P_{\text{ER-RF}},
\end{equation} where $P_{\text{ER-RF}}$ is the received RF power at the ER and $\eta \in [0, 1]$ is a fixed constant that reflects the quality of the RF-to-DC conversion circuit, i.e., the power conversion efficiency.
Eq. \eqref{eq_3} implies that the energy conversion efficiency is  independent of the input power\footnote{In this work, a normalized energy unit is assumed, i.e., Joule-per-second. In other words, the terms ``energy" and ``power" are interchangeable.} level at the ER. However, it is expected that practical EH circuits introduce non-linearity into the end-to-end WPT, cf. \cite{valenta2014harvesting}--\nocite{le2008efficient}\cite{guo2012improved}. In particular, due to limitations in practical EH circuits, the RF power conversion efficiency improves with increasing power with diminishing returns until it reaches a saturation value \cite{valenta2014harvesting}, which corresponds to the maximum possible harvested energy. In contrast, according to the linear EH model in \eqref{eq_3} assumed in the current literature, the harvested power can arbitrarily linearly increase with the input RF power. Hence, adopting the linear EH receiver model may lead to a resource allocation mismatch. In order to be able to capture the effects of practical EH circuits on the end-to-end power conversion, we adopt the practical non-linear EH model recently proposed in \cite{comm_letter:practical}:
\begin{align}
E_k(n) &= \frac{ \big[ \Psi_k(n) - M \Omega \big] }{1 - \Omega}, \,\, \Omega = \frac{1}{1+e^{ab}}, \label{eq_4} \\
\Psi_k(n) &= \frac{M}{1+e^{-a( P_{\text{ER}_k}(n) h_k(n)-b)}}. \label{eq_5}
\end{align}
Here, $\Psi_k(n)$ is the traditional logistic function with respect to the received RF power $P_{\text{ER}_k}(n)$ of user $k$ in time slot $n$, $\forall n, k$. The practical non-linear EH model can capture the joint effects of different non-linear phenomena caused by hardware constraints including circuit sensitivity limitations and current leakage \cite{le2008efficient,guo2012improved} by adjusting the parameters $a$, $b$, and $M$ \cite{comm_letter:practical}. In particular, $M$ denotes the maximum harvested power at an ER when the EH circuit is saturated, while $a$ and $b$ are related to the detailed EH circuit specifications. Parameters $M$, $a$, and $b$ can be easily obtained by standard curve fitting based on measurement data for a given EH circuit implementation. As an example, Figure~\ref{fig:curve_fit} shows that the curve fitting for the non-linear EH model in \eqref{eq_4} with parameters $M=0.024$, $b=0.0014$, and $a=1500$ closely matches experimental results provided in \cite{guo2012improved} for the wireless power harvested by a practical EH circuit. Figure~\ref{fig:curve_fit} also illustrates the inability of the linear model in \eqref{eq_3} to accurately model the characteristics of practical EH circuits over the entire range of input powers.

\vspace*{-1mm}
\section{Resource Allocation}\label{sect:resource_allocation}
In the following, we adopt the non-linear EH model in \eqref{eq_4} and study the resource allocation algorithm design for the downlink multiuser SWIPT system in Figure~\ref{fig:system_model}. Joint user scheduling and power allocation is performed assuming availability of full CSI at the base station. \footnote{CSI can be obtained in every time slot by exploiting feedback from users in frequency division duplex (FDD) systems and channel reciprocity in time division duplex (TDD) systems.} Furthermore, since $\Omega$ does not affect the design of the optimal user scheduling and power allocation, cf. \eqref{eq_4}, for notational simplicity and without loss of generality, we directly use $\Psi_k(n)$, $\forall n, k,$ from \eqref{eq_5} to represent the harvested power at the ERs.
\subsection{Optimization Problem Formulation}
\label{subsect:opt_problem_form}
The system design objective is to maximize the average total harvested power in the system for the practical non-linear EH model. Hence, we formulate the resource allocation algorithm design as the following optimization problem, with respect to the user selection and the power allocation variables, $s_k(n)$ and $P_k(n)$, $\forall n, k$, respectively:
\begin{align}
\label{eq_6}
\underset{s_k(n), P_k(n)}{\maxo} \,\, & \frac{1}{T}\sum_{n=1}^{T} \sum_{k=1}^{K} \frac{M}{1+e^{-a(P_{\text{ER}_k}(n)h_k(n)-b)}}\\
\mathrm{s. t.} \,\, & \mathrm{C1:} \ s_k(n) \in \{ 0,1 \} , \forall n, k, \nonumber \\
& \mathrm{C2:} \ \sum_{k=1}^{K} s_k(n) \leq 1 , \forall n,  \nonumber \\
& \mathrm{C3:} \ \frac{1}{T} \sum_{n=1}^{T}  \sum_{k=1}^{K} P_k (n) s_k(n) \leq P_{\text{av}} , \nonumber \\
& \mathrm{C4:} \ \sum_{k=1}^{K} P_k(n) s_k(n) \leq P_{\text{max}} ,\forall n ,  \nonumber \\
& \mathrm{C5:} \ \frac{1}{T}\sum_{n=1}^{T} C_k(n) s_k(n) \geq C_{\text{req}_{k}} , \forall k. \nonumber
\end{align}
Variable $P_{\text{ER}_k}(n) = (1-s_k(n))(\sum_{j=1}^{K}s_j(n)P_j(n)), \forall n, k,$ in the objective function is the total RF power received at ER $k$ in time slot $n$. In the considered problem, we focus on the long-term system performance for $T \rightarrow \infty$. Constraints C1 and C2 are imposed to guarantee that in each time slot $n$, at most one user is served by the transmitter for ID. C3 is a constraint on the average radiated power $P_\text{av}$,  and C4 constrains the maximum transmit power $P_{\text{max}}$ in each time slot, which may be limited because of hardware constraints. Moreover, C5 represents a quality of service (QoS) constraint, where $C_k(n)$ is the maximum achievable data rate in \eqref{eq_2} for user $k$ in time slot $n$. C5 ensures that the average data rate of user $k$ needs to satisfy the minimum required data rate $C_{\text{req}_{k}}$.




\subsection{Optimization Problem Solution} \label{sect:solution}

The objective function in \eqref{eq_6} is a sum-of-ratios function which is a non-convex function. Furthermore, the binary integer constraint $\text{C1}$ makes the optimization problem combinatorial in nature. In order to efficiently solve problem \eqref{eq_6}, we transform it into a more tractable equivalent\footnote{In this paper, two optimization problems are considered equivalent if both problems share the same solution.} optimization problem. The first step in obtaining a solution for the considered non-convex problem in \eqref{eq_6} is to transform the sum-of-ratios objective function.

\emph{Objective Function Transformation: }In general, computationally efficient algorithms, such as the Dinkelbach method \cite{JR:DinB_method} or the Charnes-Cooper transformation, can be adopted to solve non-linear optimization problems having a single-ratio objective function and a convex feasible set. However, these popular approaches cannot be applied to sum-of-ratios objective functions. The method recently introduced in \cite{jonga2012efficient}, on the other hand, offers a solution to the sum-of-ratios problem and was shown to achieve the global optimum. Following the same procedure as in \cite{jonga2012efficient}, in the following theorem, we introduce a transformation for the objective function in \eqref{eq_6}.

\begin{Thm}\label{thm:non_linear_sum_of_ratios}
Let $s_k^*(n)$ and $P_k^*(n)$ be the optimal solution to the optimization problem in \eqref{eq_6}. Then, there exist two parameters $\mu_k^*(n)$ and $\beta_k^*(n)$, $\forall n, k$, such that $s_k^*(n)$, and $P_k^*(n)$ are also the optimal solution to the following optimization problem
\begin{equation}
\hspace*{-1mm}\underset{s_k(n), P_k(n) \in \mathcal{C}}{\maxo} \frac{\hspace*{-1mm}\sum\limits_{n=1}^{T}\hspace*{-0.5mm}\sum\limits_{k=1}^{K} \hspace*{-0.8mm} \mu_k^*(n) \hspace*{-1mm}\Bigg[\hspace*{-1mm} M\hspace*{-1mm} -\hspace*{-1mm} \beta_k^*(n) \big(\hspace*{-0.5mm} 1 + e^{-a(P_{\mathrm{ER}_k}(n)h_k(n)-b)} \big)\hspace*{-1mm} \Bigg]\hspace*{-1mm}}{T} .
\label{eq_7}
\end{equation}Here, $\mathcal{C}$ is the feasible solution set of the problem in \eqref{eq_6}. In addition, the optimization variables $s_k^*(n)$ and $P_k^*(n)$ must satisfy the system of equations:
\begin{align}
\beta_k^*(n) \big( 1 + e^{-a ( P_{\mathrm{ER}_k}^*(n) h_k(n) - b)} \big) - M  &= 0, \label{eq_8} \\
\mu_k^*(n) \big( 1 + e^{-a ( P_{\mathrm{ER}_k}^*(n) h_k(n) - b)} \big) - 1  &= 0, \forall n, k. \label{eq_9}
\end{align}
\end{Thm}

\emph{Proof: }Please refer to \cite{jonga2012efficient} for a proof of Theorem \ref{thm:non_linear_sum_of_ratios}.\qed

Theorem \ref{thm:non_linear_sum_of_ratios} states that for the maximization problem with a sum-of-ratios objective function in \eqref{eq_6}, there exists an equivalent optimization problem with an objective function in parametric subtractive form, such that both problem formulations lead to the same optimal user selection and power allocation policy. As a result, we can focus on the equivalent objective function in \eqref{eq_7} in the rest of this paper. Moreover, the optimization problem can be solved efficiently by an iterative resource allocation algorithm, as will be shown in the next section.

\vspace*{-2mm}
\subsection{Iterative Resource Allocation Algorithm}

In the following, we focus on the design of a computationally efficient algorithm for achieving the globally optimal solution of the resource allocation optimization problem in \eqref{eq_6}. The algorithm consists of two nested loops. In the inner loop, we solve the optimization problem with the transformed objective function in \eqref{eq_7} for given $(\mu_k(n), \beta_k(n)), \forall n, k$. Then, in the outer loop, we find the optimal $(\mu_k^*(n), \beta_k^*(n)), \forall n, k,$ satisfying equations \eqref{eq_8} and \eqref{eq_9}, cf. Algorithm 1 in Table \ref{table:algorithm}.

\subsubsection{Solution of the Inner Loop}

Although the objective function in \eqref{eq_7} is in subtractive form and concave with respect to the optimization variables, the transformed optimization problem is still non-convex due to binary constraint $\text{C1}$ and the coupling between the optimization variables $s_k(n)$ and $P_k(n)$ in constraints C3, C4, and C5. To obtain a tractable problem formulation, we first handle the binary constraint C1 in \eqref{eq_6}/\eqref{eq_7}. For this purpose, we apply time-sharing relaxation by following a similar approach as in \cite{CN:Maryna_2015,wong1999multiuser}. In particular, we relax the user selection variables $s_k(n)$ in constraint $\text{C1}$ of \eqref{eq_6}/\eqref{eq_7} such that the variables can assume real values between $0$ and $1$, i.e., $\widetilde{\text{C1:}} \,\, 0 \leq s_k(n) \leq 1, \forall n, k$. The user selection variables can now be interpreted as time-sharing factors for the $K$ users during time slot $n$. Next, to facilitate the power allocation under time-sharing, we introduce the auxiliary variable $P_k'(n) = P_k(n)s_k(n)$, $\forall n, k,$ to the optimization problem. The new optimization variable $P_k'(n)$ represents the actual transmitted power in the RF of the transmitter for user $k$ in time slot $n$ under the time-sharing assumption. Besides, we also introduce an auxiliary optimization variable $P_k^{\text{virtual}}(n)=(1-s_k(n))\sum_{k=1}^{K}P_k'(n)$, which represents the actual received power at EH receiver $k$ in time slot $n$, to decouple the optimization variables in the objective function. Consequently, the inner loop optimization problem, which we solve in each iteration of the resource allocation algorithm, is rewritten with respect to the optimization variables $s_k(n)$, $P_k'(n)$, and $P_k^{\text{virtual}}(n)$ as:
\begin{align}\label{eq_11}
&\hspace*{-1mm}\underset{\substack{s_k(n),P_k'(n),\\P_k^{\text{virtual}}(n)}}{\maxo}\,\, \frac{\hspace*{-1mm}\sum\limits_{n=1}^{T}\hspace*{-0.5mm}\sum\limits_{k=1}^{K} \hspace*{-0.8mm} \mu_k(n) \hspace*{-1mm}\Bigg[\hspace*{-1mm} M\hspace*{-0.3mm}\hspace*{-0.5mm} -\hspace*{-0.3mm}\hspace*{-0.5mm} \beta_k(n) \big(\hspace*{-0.5mm} 1\hspace*{-0.5mm} + \hspace*{-0.5mm}e^{-a(P_k^{\mathrm{virtual}}(n) h_k(n)-b)} \big)\hspace*{-1mm} \Bigg]\hspace*{-1mm} }{T} \nonumber \\
 & \mathrm{s. t. } \, \,  \widetilde{\mathrm{C1:}}\, 0 \leq s_k(n) \leq 1 , \forall n, k , \, \, \mathrm{C2:} \sum_{k=1}^{K}  s_k(n) \leq 1 , \forall n, \nonumber \\
 & \mathrm{C3:} \frac{1}{T}  \sum_{n=1}^{T}  \sum_{k=1}^{K} P_k'(n) \leq P_{\mathrm{av}}, \, \, \hspace*{1.8mm} \mathrm{C4:} \sum_{k=1}^{K}  P_k'(n) \leq P_{\mathrm{max}}, \forall n, \nonumber \\
&\mathrm{C5:} \frac{1}{T} \sum\limits_{n=1}^{T} s_k(n) \log_2 \Big( 1+ \frac{P_k'(n)h_k(n)}{s_k(n)\sigma^2} \Big) \geq C_{\mathrm{req}_{k}}, \forall k,\nonumber \\
 & \mathrm{C6:} P_k^{\mathrm{virtual}}(n) \leq (1-s_k(n))P_{\mathrm{max}}, \forall n,k,\nonumber \\
 & \mathrm{C7:} P_k^{\mathrm{virtual}}(n) \leq \sum_{k=1}^{K} P_k'(n) ,\forall n, k,  \nonumber \\
 & \mathrm{C8:} P_k^{\mathrm{virtual}}(n) \geq 0, \forall n, k.
\end{align}
 Constraints C6\textendash C8 are introduced due to the proposed transformation including the auxiliary variable $P_k^{\text{virtual}}(n)$. These constraints guarantee that variable $P_k^{\text{virtual}}(n)$ is consistent with the original problem formulation.

Specifically, if the time-sharing relaxation is tight, i.e., $s_k(n)\in\{0, 1\}, \forall n, k$, then \eqref{eq_11} is equivalent to the optimization problem in \eqref{eq_7}. We note that the optimization problem in \eqref{eq_11} is jointly concave with respect to the optimization variables and can be efficiently solved by standard numerical methods for convex programs, such as the gradient method or the interior point method \cite{book:convex}. In other words, optimal power allocation and scheduling policies for \eqref{eq_11} can be obtained numerically.  Now, we study the tightness of the adopted time-sharing relaxation.

\begin{Thm}\label{thm:appendix}
Problems \eqref{eq_7} and \eqref{eq_11} are equivalent and have the same optimal solution, despite the time-sharing relaxation in constraint $\widetilde{\mathrm{C1}}$ of \eqref{eq_11}. In particular, the time-sharing relaxation is tight and the optimal solution of \eqref{eq_11} satisfies constraint $\text{C1}$ in \eqref{eq_7}, i.e.,  $s_k(n) \in \{0, 1\}$, $\forall n,k$.
\end{Thm}

\emph{Proof: }Please refer to the Appendix.\qed
\begin{table}[t]\caption{Iterative Resource Allocation Algorithm.}\label{table:algorithm}
\vspace*{-5mm}
\begin{algorithm} [H]                    
\caption{Iterative Resource Allocation Algorithm}          
\label{alg1}                           
\begin{algorithmic}[1]
\STATE {Initialize maximum number of iterations  $I_{\text{max}}$, iteration index $m=0$, $\mu_k(n)$ and $\beta_k(n)$, $\forall n, k$}
\REPEAT[Outer Loop]
\STATE{Solve the transformed inner loop convex optimization problem in \eqref{eq_11} for given $\mu_k^m(n)$ and $\beta_k^m(n)$ and obtain the intermediate solution for $s_k(n)$, $P_k^{\text{virtual}}(n)$, and $P_k'(n)$, $\forall n, k$}
\IF{convergence condition in \eqref{eq_8}, \eqref{eq_9} is satisfied}
\STATE {Convergence = \textbf{true}}
\RETURN optimal user selection and power allocation
\ELSE
\STATE {Update $\mu_k^m(n)$ and $\beta_k^m(n)$, $\forall n, k,$ according to the modified Newton method \eqref{eq_15}, and set $m=m+1$}
\STATE {Convergence = \textbf{false}}
\ENDIF
\UNTIL{Convergence = \textbf{true} or $m = I_{\text{max}}$}
\end{algorithmic}
\end{algorithm}
\vspace*{-8mm}
\end{table}

\noindent  As shown in the Appendix, although we consider an infinite
number of time slots and long-term averages for the total harvested energy and the sum rate in \eqref{eq_7}, the optimal multiuser power allocation and scheduling policies
depend only on the current time slot, i.e., online scheduling is optimal.

\subsubsection{Solution of the Outer Loop}

In the outer loop of the algorithm, cf. Table \ref{table:algorithm}, we find the optimal ($\mu_k^*(n)$, $\beta_k^*(n)$), $\forall n, k,$ that satisfy \eqref{eq_8} and \eqref{eq_9}. For that purpose, we implement the modified Newton method \cite{jonga2012efficient}, as shown in the following.
For notational simplicity, we introduce parameter $\bm{\rho} = [\rho_1, \ldots, \rho_{2N}] = [\mu_1,\ldots,\mu_N, \beta_1,\ldots,\beta_N]=(\bm{\mu}, \bm{\beta})$ and functions
$\varphi_i (\rho_i) =  \rho_i \big( 1 + e^{-a ( P_i^{\text{virtual}} h_i - b)} \big) - 1$, and
$\varphi_{N+i} (\rho_{N+i}) = \rho_{N+i} \big( 1 + e^{-a ( P_i^{\text{virtual}} h_i - b)} \big) - M$, where $i \in \{1,\cdots,N\}$, and $N=T K$ is the number of terms in the sum.
In \cite{jonga2012efficient}, it is proven that the optimal solution $\bm{\rho}^* = (\bm{\mu}^*, \bm{\beta}^*)$ is achieved if and only if
$\bm{\varphi}(\bm{\rho}) = [\varphi_1,\cdots,\varphi_{2N}] = \bm{0}$ is satisfied.
In the $m$-th iteration, we update $\bm{\rho}=(\bm{\mu}, \bm{\beta})$, in the following manner:
\begin{equation}
\bm{\rho}^{m+1} = \bm{\rho}^m + \zeta^m \bm{q}^m, \,\, \label{eq_15} \\
\end{equation}
where $\bm{q}^m = [\bm{\varphi}'(\bm{\rho})]^{-1}\bm{\varphi}(\bm{\rho})$, and $[\cdot]^{-1}$ denotes the inverse of a matrix. Here, $\bm{\varphi}'(\bm{\rho})$ is the Jacobian matrix of $\bm{\varphi}(\bm{\rho})$ \cite{jonga2012efficient}. Moreover, $\zeta^m$ is defined as the largest $\varepsilon^l$ that satisfies:
\begin{equation}
\label{eq_16}
\hspace*{-1mm}\norm{\bm{\varphi} (\bm{\rho}^m + \varepsilon^l \bm{q}^m)}\leq  (1-\delta \varepsilon^l) \norm{\bm{\varphi}(\bm{\rho}^m)},
\end{equation}
where $l \in  \{ 1,2,\cdots \}$, $\varepsilon^l \in (0,1)$, $\delta \in (0,1)$, and $\norm{\cdot}$ denotes the Euclidean vector norm.
It is shown in \cite[Theorem 3.3]{jonga2012efficient} that the modified Newton method converges to the unique solution $\bm{\rho}^* = (\bm{\mu}^*, \bm{\beta}^*)$ with linear rate for any starting point, while satisfying \eqref{eq_8} and \eqref{eq_9}.


\emph{Remark: }We note that the computational complexity of the proposed algorithm is polynomial time, which is considered to be fast and computationally efficient in the literature \cite[Chapter 34]{book:polynomial}. This characteristic is desirable for real-time implementation of the algorithm.



\section{Results}
\label{sect:result-discussion}

In this section, we evaluate the performance of the proposed resource allocation algorithm design for the practical non-linear EH model through computer simulations. For the specific simulation settings, we assume a carrier frequency of $915$ MHz and a signal bandwidth of $200$ kHz \cite{JR:ng2014multi}. The thermal noise power is $\sigma^2 = -120 $ dBm. The simulations are performed for $10$ and $15$ users in the system  by averaging over different channel realizations. We assume the path loss model defined in \cite{rappaport1996wireless}, with a path loss exponent of two. The multipath fading coefficients are modelled as independent and identically distributed Rician fading with Rician factor $0$ dB. The transmit antenna gain is set to $18$ dBi, while the receive antenna gain is $0$ dBi. The average radiated power $P_{\text{av}}$ is constrained to $20\%$ of the maximum transmit power $P_{\text{max}}$. For the non-linear EH model parameters, cf. \eqref{eq_4}, \eqref{eq_5}, we assume $M = 24$ mW, which corresponds to the maximum harvested power at the EH receiver. Besides, we adopt $a = 1500$ and $b = 0.0014$, which are obtained by curve fitting from the measurement data in \cite{guo2012improved}. We assume $C_{\text{req}} = 3$ bit/s/Hz for the ID users. Extensive simulations (not shown here) have revealed that, in general, the proposed iterative resource allocation algorithm converges to the globally optimal solution after less than $20$ iterations.
\begin{figure}
\centering
\includegraphics[width= 3.5in]{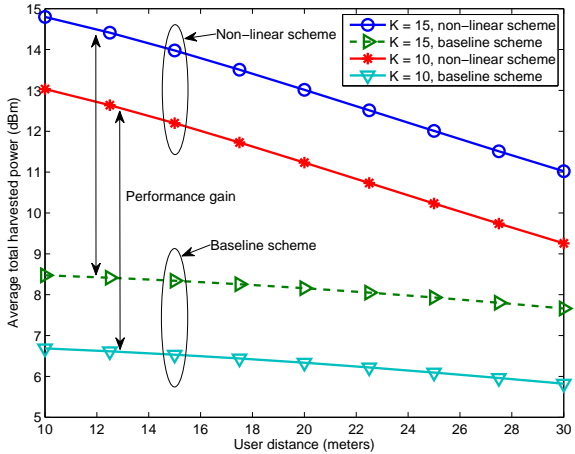}
\vspace*{-1mm}
\caption{Average total harvested power (dBm) versus user distances (meters) for different numbers of users.}
\vspace*{-2mm}
\label{fig:HP_distances}
\end{figure}

Figure~\ref{fig:HP_distances} depicts the average total harvested power versus the distance of the users from the base station for the proposed resource allocation algorithm. The maximum transmitted power in every time instant was chosen to be $P_{\text{max}} = 46$ dBm. Furthermore, for simplicity, we assume equal distances between the base station and all users, such that all users have the same channel gain-to-noise ratio.
Figure~\ref{fig:HP_distances} shows that the average total harvested power is a decreasing function with respect to the distance of the users from the base station. This is mainly because of the pronounced reduction of the power density of the received RF signal for increasing user distance from the base station. Besides, the RF-to-DC conversion efficiency of the EH circuit degrades significantly at lower received RF power values due to sensitivity limitations \cite{valenta2014harvesting}. On the other hand, the total harvested power increases when there are more users in the system, since a larger portion of the radiated power can be harvested. For comparison, we also plot the performance of a baseline scheme\footnote{ We note that the results in \cite{multiuser_scheduling, CN:Maryna_2015} are not used for comparison as the objective in \cite{multiuser_scheduling, CN:Maryna_2015} was the maximization of the users' sum rate under EH constraints assuming the linear EH model, which differs from the objective in this paper.} in Figure~\ref{fig:HP_distances}. The  baseline scheme maximizes the total harvested power assuming the conventional linear model in \eqref{eq_3} subject to the constraint set in \eqref{eq_6}, i.e., the EH model assumed for optimization is not matched to the practical non-linear model adopted in the simulation. The conversion efficiency for the linear model was chosen to be $\eta = 0.5$ \cite{multiuser_scheduling,CN:Maryna_2015}. Due to the resulting resource allocation mismatch, the baseline scheme results in an evidently smaller amount of total harvested energy compared to the proposed scheme.

\begin{figure}
\centering
\includegraphics[width= 3.5in]{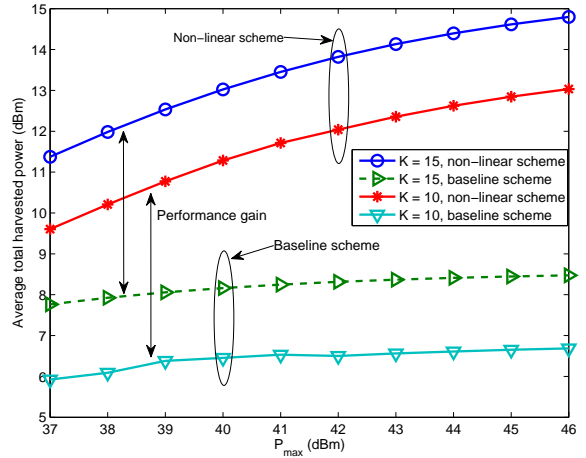}
\vspace*{-1mm}
\caption{Average total harvested power (dBm) versus the maximum transmitted power in each time slot, $P_{\text{max}}$ (dBm), for different numbers of users.}
\vspace*{-3mm}
\label{fig:HP_p_max}
\end{figure}
In Figure~\ref{fig:HP_p_max}, we show the average total harvested power versus the maximum transmit power allowance $P_{\text{max}}$. We assume that the distance between the base station and each user is $10$ meters. The average total harvested power is an increasing function with respect to $P_{\text{max}}$ for the proposed resource allocation algorithm optimized for the non-linear EH model. This increasing trend continues until the maximum possible power is harvested at all EH receivers, i.e., all EH circuits are saturated. On the other hand, the average total harvested power is almost constant with respect to $P_{\text{max}}$ for the baseline scheme, which was again optimized for the linear EH model. In fact, the baseline scheme may cause saturation in some EH receivers and underutilization of other EH receivers because of the characteristics of the non-linear EH circuits. For instance, the linear scheme allocates an exceedingly large amount of resources to the user with the best channel conditions for EH. In contrast, the proposed resource allocation algorithm optimized for the non-linear model distributes the available power more evenly across EH receivers and across time in order to avoid saturation and underutilization.
\section{Conclusions}\label{sect:conclusion}
In this paper, we designed a joint scheduling and power allocation
algorithm for the maximization of the long-term average total harvested power in a multiuser SWIPT system, where a practical non-linear EH model for the end-to-end WPT was adopted. Simulation results revealed that adopting a realistic non-linear EH model instead of the conventional linear model for resource allocation algorithm design may substantially increase the performance of SWIPT systems employing practical EH circuits.
\section*{Appendix - Proof of Theorem 2}\label{app:time_sharing}
In this section, we follow a similar approach as in \cite{wong1999multiuser} to prove Theorem \ref{thm:appendix}. First, we introduce the Lagrangian for \eqref{eq_11}:
\begin{align}
\label{eq_a_01}
&\mathcal{L}(P_k^{\text{virtual}}(n), P_k'(n), s_k(n), \mu_k(n), \beta_k(n), \mathcal{D}) \\
&= \sum_{n=1}^{T} \sum_{k=1}^{K} \mu_k(n) \Big( M-\beta_k(n)\big( 1+e^{-a( P_k^{\text{virtual}}(n)h_k(n)- b)}\big)\Big) \nonumber \\ &- \sum_{n=1}^{T}\lambda(n)\Big( \sum_{k=1}^{K} s_k(n)-1\Big) -\sum_{n=1}^{T} \sum_{k=1}^{K}\alpha_k(n)\Big( s_k(n)-1\Big) \nonumber \\&+\sum_{n=1}^{T} \sum_{k=1}^{K}\vartheta_k(n)s_k(n)-\gamma \Big( \frac{1}{T} \sum_{n=1}^{T} \sum_{k=1}^{K}P_k'(n) - P_{\text{av}} \Big) \nonumber \\
&-\sum_{n=1}^{T}\varrho(n)\Big( \sum_{k=1}^{K}P_k'(n)-P_{\text{max}}\Big)+\sum_{n=1}^{T} \sum_{k=1}^{K}\theta_k(n)P_k^{\text{virtual}}(n)\nonumber \\
&-\sum_{k=1}^{K} \epsilon(k)\Big( C_{\text{req}_{k}}-\frac{1}{T}\sum_{n=1}^{T} s_k(n)\log_2\big(1+\frac{P_k'(n)h_k(n)}{s_k(n)\sigma^2}\big)\Big)\nonumber \\
&-\sum_{n=1}^{T} \sum_{k=1}^{K}\zeta_k(n)\Big( P_k^{\text{virtual}}(n)-\big(1-s_k(n)\big)P_{\text{max}}\Big) \nonumber \\ &-\sum_{n=1}^{T} \sum_{k=1}^{K}\eta_k(n)\Big( P_k^{\text{virtual}}(n)-\sum_{k=1}^{K} P_k'(n)\Big), \nonumber
\end{align}
where $\mathcal{D} = \{\alpha_k(n)$, $\vartheta_k(n)$, $\lambda(n)$, $\gamma$, $\varrho(n)$, $\epsilon(k)$, $\zeta_k(n)$, $\eta_k(n)$, $\theta_k(n)\}$, $\forall n, k,$ is the set containing all Lagrange multipliers for constraints $\widetilde{\text{C1}}$--C8 in \eqref{eq_11}. After differentiating the Lagrangian in \eqref{eq_a_01} with respect to $s_k(n)$, we obtain:
\begin{align}
\label{eq_a_04}
\hspace*{-2mm}\frac{\partial \mathcal{L}}{\partial s_k(n)}\hspace*{-1mm} &= \hspace*{-1mm}\frac{\epsilon(k)F_k(n)}{T \ln 2} \hspace*{-1mm}-\hspace*{-1mm}\lambda(n)\hspace*{-1mm} +\hspace*{-1mm} g_k(n) \hspace*{-1mm} \begin{cases}
  < \hspace*{-1mm}0, & \hspace*{-3mm}\text{if } s_{k^*}(n)\hspace*{-1mm}=\hspace*{-1mm}0, \\
  =\hspace*{-1mm}0, & \hspace*{-3mm}\text{if } s_{k^*}(n) \hspace*{-1mm}\in \hspace*{-1mm}(0,1),\\
  > \hspace*{-1mm}0, & \hspace*{-3mm}\text{if } s_{k^*}(n)\hspace*{-1mm}=\hspace*{-1mm}1,
\end{cases}
\end{align}
where $g_k(n)= \vartheta_{k}^*(n) - \alpha_{k}^*(n) -\zeta_{k}^*(n)P_{\text{max}}$, and $F_k(n)=\Big[\ln \big( 1\hspace*{-1mm}+\hspace*{-1mm}\frac{P_k(n)h_k(n)}{\sigma^2}\big)\hspace*{-1mm} -\hspace*{-1mm} \frac{\frac{P_k(n)h_k(n)}{\sigma^2}}{1+\frac{P_k(n)h_k(n)}{\sigma^2}} \Big]$, $\forall n, k$. The dual variables $\alpha_k(n),\vartheta_k(n),\zeta_k(n)$, and $\epsilon(k)$ are considered as constants, and their optimal values $\alpha_k^*(n),\vartheta_k^*(n),\zeta_k^*(n)$, and $\epsilon^*(k)$ can be obtained in an offline manner.
Following the proof and conditions for the derivatives given in \cite[Section IV]{wong1999multiuser}, the optimal user selection criterion reduces to:
\begin{align}
\label{eq_a_05}
s_{k^*}(n) = \begin{cases}
  0, & \text{if } \lambda(n) > \frac{\epsilon(k)}{T \ln 2} F_k(n) + g_k(n) , \\
  1, & \text{if } \lambda(n) < \frac{\epsilon(k)}{T \ln 2} F_k(n) + g_k(n).
\end{cases}
\end{align}
If $F_k(n)$ is different for every user $k$ and time slot $n$, the optimal scheduling policy is given by:
\begin{align}
\label{eq_a_06}
s_{k^*}(n) = 1 , \,\, \text{and } s_{k}(n) = 0, \forall k \neq k^*, \forall n,
\end{align}
where $k^*\hspace*{-1mm}=\hspace*{-1mm}\text{arg}\max_{k} \Big(\hspace*{-0.5mm} \frac{\epsilon(k)}{T \ln 2} F_k(n) + g_k(n) \hspace*{-0.5mm}\Big)$ denotes the optimal user selection index for ID in time slot $n$. We note that the Lagrange multipliers in the scheduling policy  depend only on the statistics of the
channels. Hence, they can be calculated offline, e.g. using the gradient method, and then be used for online scheduling as long
as the channel statistics remain unchanged. As a result, the optimal scheduling rule in \eqref{eq_a_05} depends only on the CSI in the current time slot and the channel statistics, i.e., online scheduling is optimal, although the considered optimization problem in \eqref{eq_7} considers an infinite number of time slots and long-term averages for the total harvested energy. Furthermore, the solution of the relaxed problem given in \eqref{eq_a_05} is Boolean in nature. Therefore, the proposed time-sharing relaxation in constraint $\widetilde{\text{C1}}$ in \eqref{eq_11} is tight. \qed

\end{document}